\documentclass[prc,unsortedaddress,groupedaddress,reprint,amsmath,amsfonts,amssymb,showpacs,floatfix,nofootinbib]{revtex4-1}
\usepackage{epsfig}
\usepackage{multirow}
\usepackage{relsize}
\usepackage{longtable}
\usepackage{color}
\usepackage{graphicx}
\begin{document}

\title{Subthreshold resonances and resonances in the $R$-matrix method for binary reactions and in the Trojan Horse method}

\author{A. M. Mukhamedzhanov$^{1}$}
\email{akram@comp.tamu.edu} 
\author{Shubhchintak$^{2}$}
\email{shub.shubhchintak@tamuc.edu} 
\author{C. A.  Bertulani$^{2,3}$}
\email{carlos.bertulani@tamuc.edu}

\affiliation{$^{1}$Cyclotron Institute, Texas A$\&$M University, College Station, TX 77843, USA}
\affiliation{$^{2}$Department of Physics and Astronomy, Texas A$\&$M University-Commerce, Commerce, TX 75429, USA}
\affiliation{$^{3}$Department of Physics and Astronomy, Texas A$\&$M University, College Station, TX 77843, USA}

\date{\today}

\begin{abstract}
In this paper we discuss the $R$-matrix approach to treat the subthreshold resonances for the single-level and one channel,  and for the single-level and two channel cases. 
In particular, the expression relating the ANC with the observable reduced width, when the subthreshold bound state is the only channel or coupled with an open channel, which is a resonance, is formulated. Since the ANC plays a very important role in nuclear astrophysics, these relations significantly enhance the power of the derived equations.  
We present the relationship between the resonance width and the ANC for the general case and consider two limiting cases: wide and narrow resonances. 
Different equations for the astrophysical S-factors in the $R$-matrix approach are presented.  After that we discuss the Trojan Horse Method (THM) formalism. The developed equations are obtained using the surface-integral formalism and the generalized $R$-matrix approach for the three-body resonant reactions.  It is shown how the Trojan Horse (TH) double differential cross section can be expressed in terms of the on-the-energy-shell  astrophysical S-factor for the binary sub-reaction. Finally, we demonstrate how the THM can be used to calculate the astrophysical S-factor for the neutron generator $^{13}{\rm C}(\alpha,\,n)^{16}{\rm O}$ in low-mass AGB stars.  At astrophysically relevant energies this astrophysical S-factor is controlled by the threshold level $1/2^{+}, E_{x}= 6356$ keV.   Here, we reanalyzed  recent TH data taking into account more accurately the three-body effects and using both assumptions that the threshold level is a subthreshold bound state or  it is a resonance state.
\end{abstract}


\maketitle

\section{Introduction}

A subthreshold bound state (which is close to threshold) reveals itself as a subthreshold resonance in low-energy scattering or reactions. Subthreshold resonances play an important role in low energy processes, in particular, in astrophysical reactions.  In this paper we present new $R$-matrix equations for the reaction amplitudes and astrophysical $S$-factors for analysis of  reactions proceeding through  subthreshold resonances. We consider elastic scattering and resonant reaction for subthreshold resonance coupled with open resonance channels for single and two-level cases.  All the equations are expressed in terms of the formal and observable reduced widths.  The observable  reduced width is expressed in terms of the asymptotic normalization coefficient (ANC). As a new result, we obtain a new equation for the connection of the ANC with the observable reduced width of the subthreshold resonance, which is coupled with a resonance channel.  This equation is extremely important taking into account a crucial role of the ANC in nuclear astrophysics.  
We also derive equations for the  Trojan Horse Method (THM) reaction amplitude, triple and double differential cross sections in the presence of the subthreshold state using a generalized $R$-matrix method and the surface-integral method. The THM is a powerful indirect technique to treat resonances. We show that the THM can equally well incorporate the subthreshold and real resonances. 

The Trojan Horse (TH) double differential cross section is expressed in terms of the  on-the-energy-shell (OES) astrophysical factor, which can be contributed by both the subthreshold and resonance states.
It is also demonstrated how the developed  theory can be used to calculate the astrophysical S-factor of the reaction ${}^{13}{\rm C}(\alpha,\,n){}^{16}{\rm O}$, which is a neutron generator for $s$-processes in low mass AGB stars. A special attention is given to the subthreshold $-3$ keV bound state in ${}^{17}{\rm O}= (\alpha\,{}^{13}{\rm C})$, which behaves as a subthreshold resonance. There are few papers where S-factors have been calculated for ${}^{13}{\rm C}(\alpha,\,n){}^{16}{\rm O}$, but  the equations were not described. Our theory, especially the new Trojan Horse (TH) equations, provide a very useful tool for experimentalists to treat  resonances. 

The paper is organized as follows. In Section II first we consider the single-channel, single-level elastic scattering just to demonstrate how to relate the ANC and the reduced width for the subthreshold resonance. After that the two-channel case is introduced, in which the subthreshold bound state is coupled with an open channel. The connection between the ANC and the reduced width obtained for the single-channel case is generalized for the two-channel case, when one of the channels is  closed and the second one is open. In Section III we consider the resonant reaction proceeding through the subthreshold state.  Generalization of the standard $R$-matrix equations for the reaction amplitudes is presented. We also give the explicit $R$-matrix equations for astrophysical factors obtained for different cases, which  we need to analyze the 
 ${}^{13}{\rm C}(\alpha,\,n){}^{16}{\rm O}$ reaction. To follow the results obtained in Sections II and III the reader is supposed to be familiar with the classical $R$-matrix review \cite{lanethomas58}. 
In Section IV we derive the reaction amplitude, triple and double differential cross section for the indirect TH  reactions proceeding through the intermediate resonances within the framework of the generalized $R$-matrix approach.  Finally, in Section V  we present the analysis of the astrophysical factor for the ${}^{13}{\rm C}(\alpha,\,n){}^{16}{\rm O}$ reaction, which is the neutron generator in the AGB stars.  
Throughout the paper the system of units in which $\hbar=c=1$ is used.

\section{Elastic scattering}

\subsection{Single-channel, single-level }

First we consider  the single-level, single-channel $R$-matrix approach in the presence of the subthreshold bound state (also called subthreshold resonance). 
The resonant elastic scattering amplitude in the channel $i=x+A$ with the partial wave $l_{i}$ can be written in the standard $R$-matrix form
\cite{lanethomas58}:
\begin{eqnarray}
T_{ii} = -2\,i\,e^{-2\,i\,\delta_{i}^{hs}}\,\frac{ P_{i}\,(\gamma_{i}^{(s)})^{2} }{E_{1} - E_{i} -  [S_{i}(E_{i})   - B_{i} + i\,P_{i}]\,(\gamma_{i}^{(s)})^{2} }.\nonumber\\
\label{resTii1}
\end{eqnarray}
Here, $\,\gamma_{i}^{(s)}$ is the reduced width amplitude of the subthreshold bound state $F^{s}=(x\,A)^{(s)}$ with the binding energy $\,\varepsilon_{i}^{(s)} = m_{x} + m_{A} - m_{F^{(s)}}$, $\,m_{j}$ is the mass of the particle $j$, $\,E_{i} \equiv E_{xA}$ is the $x-A$ relative kinetic energy, $E_{1}$ is the $R$-matrix energy level,  $\,S_{i}(E_{i})  = R_{i} \,Re \Big[{\rm d}\,{\rm ln}O_{l_{i}}(k_{i},r_{i})/{\rm d}r_{i}\Big|_{r_{i}=R_{i}}\Big]$ is the $R$-matrix shift function in channel $i$, $O_{l_{i}}$ is outgoing spherical wave, $r_{i} \equiv r_{xA}$ is the radius connecting centers-off-mass of the particles in the channel $i$ and $k_{i} \equiv k_{xA}$ is the $\,x-A$ relative momentum.  $\,B_{i} \equiv B_{l_{i}}$ is the energy-independent $R$-matrix boundary condition constant,  $\,P_{i}\equiv P_{l_{i}}(E_{i},\,R_{i})$ and  $\,R_{i} \equiv R_{xA}$ are the penetrability factor and the channel radius in the channel $i$,  $\,\delta_{i}^{hs} \equiv \delta_{xA\,l_{i}}^{hs}$   is the hard-sphere scattering phase shift in the channel $i$.

If we choose the boundary condition parameter $\,B_{i}= S_{i}(- \varepsilon_{i}^{(s) } )$,  in the low-energy region  where the linear approximation is valid
\begin{eqnarray}
S_{i}(E_{i}) - S_{i}(-\varepsilon_{i}^{(s)})  \approx \frac{ {\rm d}\,S_{i}(E_{i}) }{{\rm d}E_{i}}\,\Big|_{E_{i}=-\varepsilon_{i}^{(s)}} (E_i + \varepsilon_{i}^{(s)}).
\label{linearapproxDelta1}
\end{eqnarray}
Then at small $E_{i}$
\begin{eqnarray}
T_{ii} = \frac{2\,i\,e^{-2\,i\,\delta_{i}^{hs}}\, P_{i}({\tilde \gamma}_{i}^{(s)})^{2} }{\varepsilon_{i}^{(s)} + E_{i} +  i\,P_{i}\,({\tilde \gamma}_{i}^{(s)})^{2} },
\label{Tresii2}
\end{eqnarray}
which has a pole at  $\,E_{i}= -\varepsilon_{i}^{(s)}$    because $P_{i}$ vanishes for $E_{i} \leq 0$.   Here,  ${\tilde \gamma}_{i}^{(s)}$ is the observable reduced width of the subthreshold resonance. The observable reduced width $({\tilde \gamma}_{i}^{(s)})^{2}$  is related to the formal $R$-matrix  reduced width $ ( \gamma_{i}^{(s)})^{2}$  as
\begin{eqnarray}
({\tilde \gamma}_{i}^{(s)})^{2}= \frac{  ( \gamma_{i}^{(s)})^{2} }{ 1+(\gamma_{i}^{(s)})^{2} [{\rm d}S_{i}(E_{i})/{\rm d}E_{i}]\big|_{E_{i}=-\varepsilon_{i}^{(s)}}}.
\label{effredwidth1}
\end{eqnarray}

Determining the asymptotic normalization coefficient (ANC) as the residue in the pole of the scattering amplitude corresponding to the bound-state pole \cite{muk99}, we get for the ANC of the subthreshold state,
\begin{eqnarray}  
[C_{i}^{(s)}]^{2}=   \frac{2\,\mu_{i}\,R_{i}\, ({\tilde \gamma}_{i}^{(s)})^{2} }{W_{-\,\eta_{i}^{(s)},\,l_{i}+1/2}^{2}(  2\,\kappa_{i}^{(s)}\,R_{i})},
\label{ANCredwidth1}
\end{eqnarray}
where   $W_{-i\,\eta_{i}^{(s)},\,l_{i}+1/2}(\,2\,\kappa_{i}^{(s)}\,R_{i})$ is the Whittaker function, $\eta_{i}^{(s)}= (Z_{x}\,Z_{A}/137)\mu_{i}/\kappa_{i}^{(s)}$  and $\kappa_{i}^{(s)}$ are the $x-A$ Coulomb parameter and the bound-state wave number of the subthreshold state $F^{(s)}$,  $\,\mu_{i}$ is the reduced mass of $x$ and $A$,  $\,Z_{j}\,e$ is the charge of nucleus $j$.
We established the relationship between the ANC and the observable reduced width earlier in \cite{muk99}, but in the next section Eq. (\ref{ANCredwidth1}) will be generalized for two-channel case.

The observable partial resonance width of the subthreshold resonance is given by 
\begin{align}
{\tilde \Gamma}_{i}^{(s)}(E_{i}) &= 2\,P_{i}\,({\tilde \gamma}_{i}^{(s)})^{2} \nonumber\\
&= P_{i}\,\frac{\Big(C_{i}^{(s)}\,W_{-\,\eta_{i}^{(s)}\,l_{i}+1/2}(2\,\kappa_{i}^{(s)}\,R_{i}) \Big)^{2}}{\mu_{i}\,R_{i}}.
\label{GammaANC1}
\end{align}
Eq. (\ref{GammaANC1}) is of  a fundamental importance. It shows that the subthreshold bound state at $E_{i}$ behaves as a resonance with the resonance width expressed in terms of  ANC and the Whittaker function of this bound state taken on the border $R_{i}$. 
We have considered the connection between the ANC and the reduced width amplitude for the bound states.
In the next section we consider the connection between the ANC and the observable resonance width
for real resonances.

\subsection{Two-channel, single-level}  

Now  we consider the elastic scattering $x+ A \to x+A$ in the presence of the subthreshold bound state $F^{(s)}$ in the channel $i =x+ A$ which is coupled with  the second channel $f=b+B$. The relative kinetic energies in the channels $i$, $E_{i} \equiv E_{xA}$, and channel $f$,  $E_{f} \equiv E_{bB}$,  are related by 
\begin{align}
E_{f} \equiv E_{bB} = E_{i} + Q,  \qquad Q= m_{x} + m_{A} - m_{b} - m_{B} > 0.
\label{EfEiQ1}
\end{align}
We assume that $\,Q>0$, that is, the channel $f$ is open for $E_{i} \geq 0$.        
The resonance part of the elastic scattering amplitude in the channel $i=x+A$ in the single-level, two-channel $R$ matrix approach, is
\begin{eqnarray}
T_{ii} &=& -2\,i\,e^{-2\,i\,\delta_{i}^{hs}}\nonumber\\
&\times& \frac{ P_{i}\,(\gamma_{i}^{(s)})^{2} }{  E_{1} - E_{i} -  \sum\limits_{n=i,f}\,[S_{n}(E_{n}) - B_{n} + i\,P_{n}]\,\gamma_{n}^{2}},
\label{2chTii1}
\end{eqnarray}
where $\gamma_{n}$ is the formal reduced width in the channel $n=i,\,f$. Note that $\gamma_{i}  \equiv \gamma_{i}^{(s)}$. 
 $\,P_{n} \equiv P_{l_{n}} (E_{n},\,R_{n})$ and $\,R_{n}$  are   the penetrability factor and the channel radius in the channel $n$.  There are two fitting parameters, $\gamma_{i}^{(s)}$  and $\gamma_{f}$, in the single-level, two-channel $R$-matrix fit at fixed channel radii.

Again, we use the boundary condition 
$B_{n}= S_{n}(-\varepsilon_{i}^{(s)})$ and $E_{1}=-\varepsilon_{i}^{(s)}$. 
The energy $E_{i}=-\varepsilon_{i}^{(s)}$ in the channel $i$ and corresponds to $E_{f}= Q- \varepsilon_{i}^{(s)}$ in the channel $f$. Assuming a linear energy dependence of $S_{n}(E_{n})$ at small $E_{i}$, we get
\begin{eqnarray}
T_{ii} \approx 2\,i\,e^{-2\,i\,\delta_{i}^{hs}}\,\frac{P_{i}\,\big({\tilde\gamma}_{i}^{(s)}\big)^{2} }{\varepsilon_{i}^{(s)} + E_{i}  +   i\sum\limits_{n=i,f}\,P_{n}\,{\tilde \gamma}_{n}^{2}}, 
\label{2chTii12} 
\end{eqnarray}                 
where the observable reduced width  in the channel $n$ is 
\begin{eqnarray}
{\tilde \gamma}_{n }^{2}= \frac{   \gamma_{n}^{2} }{  1+\sum\limits_{t=i,f}\,\gamma_{t }^{2}[{\rm d}S_{t}(E_{t})/{\rm d}E_{t}]\big|_{E_{t}=E_{t}^{(s) } }},
\label{tildegammapr1a}
\end{eqnarray}
again noticing that $E_{i}^{(s)}= -\varepsilon_{i}^{(s)}$ and $E_{f}^{(s)}=Q - \varepsilon_{i}^{(s)}$.
Correspondingly, the observable partial resonance width in the channel $n$ is
\begin{eqnarray}
{\tilde \Gamma}_{n}(E_{n}) = 2\,P_{n}\,{\tilde \gamma}_{n}^{2},
\label{Gamman1}
\end{eqnarray}
with the total width ${\tilde \Gamma}(E_{i}) = {\tilde \Gamma}_{i}^{(s)}(E_{i}) + {\tilde \Gamma}_{f}(E_{f})$.  

The presence of the open channel coupled to the elastic scattering channel generates an additional term $\,n=f$ in the denominators of Eqs. (\ref{2chTii1}), (\ref{2chTii12}) and (\ref{tildegammapr1a}). The resonant elastic scattering amplitude in channel $f$ in the presence of the subthreshold bound state channel $i$ can be obtained from Eq. (\ref{2chTii1}) by replacing $i \leftrightarrow f$. 

Although  the scattering amplitude vanishes at $E_{i}=0$ it can be extrapolated  to the bound-state pole bypassing $E_{i}=0$ using
\begin{eqnarray}
T_{ii}  &\stackrel{E_{i} \to -\varepsilon_{i}^{(s)}}{\approx}&   2\,\kappa_{i}^{(s)}\,(-1)^{l_{i}+1}\,e^{i\,\pi\,\eta_{i}^{(s)}}\nonumber\\
&\times& \frac{R_{i}}{W_{-\eta_{i},l_{i}+1/2}^{2}(2\,\kappa_{i}^{(s)}\,R_{i})}\,
\frac{ ( {\tilde \gamma}_{i}^{(s)} )^{2}  }{ \varepsilon_{i}^{(s)} + E_{i} + i\,P_{f}\,{\tilde \gamma}_{f}^{2} }, \nonumber\\
\label{Tiipoleextrap2ch2}
\end{eqnarray}
where $\,P_{f} =P_{f}(Q- \varepsilon_{i}^{(s)},R_{f}),\,$  $\,\eta_{i}$  is the Coulomb parameter in the channel $i$.
Again, the ANC, as a residue in the pole of the scattering amplitude, is related to the observable reduced width $( {\tilde \gamma}_{i}^{ (s)} )^{2}$ of the subthreshold state by Eq. (\ref{ANCredwidth1}), in which 
now $({\tilde \gamma}_{i}^{(s)})^{2}$ is determined by 
\begin{eqnarray}
({\tilde \gamma}_{i}^{(s)})^{2}= \frac{ (\gamma_{i}^{(s)})^{2} }{  1+\sum\limits_{t=i,f}\,\gamma_{t }^{2}[{\rm d}S_{t}(E)/{\rm d}E]\big|_{E=E^{(s) } }}.
\label{tildegammapr1}
\end{eqnarray}
The derivation of the connection between the ANC and the reduced width of the subthreshold resonance  in the presence of an open channel $f$ is a generalization of Eq. (\ref{ANCredwidth1}) and is one of the main results obtained in this paper.

It follows from Eq. (\ref{Tiipoleextrap2ch2})  that in the presence of the open channel $f$ coupled with the channel $i$, the elastic scattering amplitude has the bound-state pole  shifted into the $E_{i}$ complex plane, i.e.,
$E_{i}^{(p)} = - \varepsilon_{i}^{(s)}  - i\,P_{f}(Q- \varepsilon_{i}^{(s)},R_{f})\,({\tilde \gamma}_{f})^{2}$.

We have established a connection between the ANC, the observable reduced width and the observable resonance width (at $E_{i} >0$) of the subthreshold resonance state. But, besides the subthreshold bound state in the channel $i$, in Eq. (\ref{2chTii12}) we have also a real resonance in the channel $f$. 
In \cite{muk99,izvAN,ANC} the definition of the ANC was also extended for real resonances.
 Here we remind the connection between the resonance width, the ANC, and the reduced width for the real resonance in the channel $f$ whose real part of the complex resonance energy is located at
 $E_{f}^{(R)}$.  The ANC for the resonance state is determined as the amplitude of the outgoing resonance wave \cite{ANC}, which is the generalization of the ANC definition for the bound state.
Then the general expression connecting the observable resonance width and the ANC is \cite{izvAN}
\begin{align}
C_{f}^{2}= 2\,(-1)^{l_{f}}\,\frac{k_{f}^{(0)}\,\rho \,\, e^{i[2\,\delta_{l_{f}}(k_{f}^{(0)}) -{\arctan(\rho)}/{2}]}}{(1+\rho^{2})^{1/4} + (1+ \rho^{2})^{-1/4}},
\label{ANCGammawideres1}
\end{align}
where $\rho= {{\tilde \Gamma}_{f}(E_{f}^{(R)})}/{(2E^{R}_{f})}$,
\begin{align}
k_{f}^{(0)}&=\sqrt{2\,\mu_{f}(E_{f}^{(R)}- i\,{\tilde \Gamma}_{f}(E_{f}^{(R)})/2}         \nonumber\\
&= k_{f}^{(R)}- i\,k_{f}^{(I)} 
\label{krERGR1}
\end{align}
is the complex resonant-state momentum in the channel $f$, ${\tilde \Gamma}_{f}(E_{f}^{(R)})$ is the observable resonance width at the resonance energy. $\delta_{l_{f}}(k_{f}^{(0)})$ is the potential 
(non-resonant) scattering phase shift in the channel $f$ taken at the complex resonant momentum $k_{f}^{(0)}$.
Thus, if the resonance is not Breit-Wigner type \Big($E_{f}^{(R)} >> {\tilde \Gamma}_{f}(E_{f}^{(R)})/2 \Big)$,
then to calculate the ANC from the resonance width one needs to calculate the non-resonant 
scattering phase shift at the complex energy, which is quite far from the real $k_{f}$-axis. It makes the 
quantity $\delta_{l_{f}}(k_{f}^{(0)})$ and, correspondingly, the ANC extremely model-dependent. Hence, it 
does not make sense to use the ANC for non-Breit-Wigner, i.e., for broad resonances.    
However, for narrow resonances from Eq. (\ref{ANCGammawideres1}) we get \cite{muk99,izvAN}
\begin{align}
C_{f}^{2}\approx \,\frac{\mu_{f}\,{\tilde \Gamma}_{f}(E_{f}^{(R)})}{k_{f}^{(R)}}\,
e^{i[2\,\delta_{l_{f}}(k_{f}^{(R)})- i\,\pi\,l_{f}]}.
\label{ANCGammanawrres1}
\end{align}

\section{Resonant reactions}

We present in this section  equations for the  reaction amplitudes proceeding through the subthreshold resonance  with the standard $R$-matrix equations generalized for the subthreshold state. Based on these amplitudes we obtain the corresponding  astrophysical S-factors which can be used  to analyze the experimental data obtained from direct and indirect measurements.  Note that here the expressions for the astrophysical factors are written in the convenient $R$-matrix form and  can be used by experimentalists for the analysis  of  similar reactions proceeding through the subthreshold resonance.

Let us consider the resonant reaction 
\begin{equation}
x + A \to F^{(s)} \to b +B
\label{reaction1}
\end{equation}
with $Q>0$, proceeding through an intermediate resonance,  which is a resonance in the exit channel $f$ and  the subthreshold bound state  $F^{(s)}= (x\,A)^{(s)}$ in the initial channel $i$.  We assume also that $Q- \varepsilon_{i}^{(s)} >0$, that is, the channel $f$ is open at the subthreshold bound-state pole is in the channel $i$.
 
The single-level, two-channel $R$-matrix amplitude describing  the resonant reaction in which in the initial state the colliding particles $x$ and $A$ have a subthreshold bound state and the resonance in the final channel $f=b+B$, can be obtained by generalizing the corresponding equations from  Refs. \cite{lanethomas58, barker2000}:
\begin{eqnarray}
T_{f\,i}&=&  2\,i\,e^{-\,i(\delta_{i}^{hs} +  \delta_{f}^{hs} ) }\nonumber\\
&\times& \frac{  \sqrt{  P_{f}  }\,\gamma_{f}\, \sqrt{ P_{i} }\,\gamma_{i }^{(s)} }{\varepsilon_{i}^{(s)} + E_{i} +\sum\limits_{n=i,f}\,[S_{n}(E_{n}) - S_{n}(E_{n}^{(s)}) + i\,P_{n}]\,\gamma_{n}^{2}}. \nonumber\\
\label{RmatrRes1}
\end{eqnarray}
Here we remind that $E_{i}^{(s)}= -\varepsilon^{(s)}$ and $E_{f}^{(s)}= Q - \varepsilon^{(s)}$. 
The astrophysical factor $S(E_{i})$  is given by
\begin{widetext}
\begin{eqnarray}
S(E_{i})({\rm keV.b})= \,\frac{{\hat J}_{F^{(s)}}}{{\hat J}_{x}\,{\hat J}_{A}}\,\nu_{N}^{2}\,\,E_{N}^{2}\,e^{2\,\pi\,\eta_{i}}\,\frac{20\,\pi}{\mu_{i}}\frac{ P_{f}\,P_{i}\,\gamma_{f}^{2}\,(\gamma_{i}^{(s)})^{2} }{\Big(\varepsilon_{i}^{(s)}  +E_{i} + \sum\limits_{n=i,f}\,[S_{n}(E_{n}) - S_{n}(E_{n}^{(s)})]\,{\gamma}_{n}^{2}\Big)^{2} + \Big[\sum\limits_{n=i,f}\,P_{n}\,{\gamma}_{n}^{2}\Big] ^{2}} ,
\label{Sfactr1l2ch1}
\end{eqnarray}
\end{widetext}
where  $J_{F^{(s)}}$ is the spin of the subthreshold state in the channel $i=x+A$, which is also the spin of the resonance in the channel $f=b+B$, $\,J_{j}$ is the spin of the particle $j$,   ${\hat J}= 2\,J+1$,   $\,\nu_{N}=0.2118$ fm is the nucleon Compton wavelength,  $E_{N}=931.5$ MeV is the atomic unit mass. All the reduced width amplitudes are expressed in MeV$^{1/2}$.

Assume now that the low-energy binary reaction (\ref{reaction1}) is contributed by a few non-interfering levels. The subthreshold resonance in the channel $\,i=x+A$, which is coupled to the open channel $f=b+B$ is attributed to the first level, $\lambda=1$, while other levels with $\lambda >1$ are attributed to two open coupled channels  $i$ and $f$ of higher energy levels  $E_{\lambda}$ and  spins $J_{F^{\lambda}}$.
The astrophysical factor $S(E_{i})$ is given by
\begin{align}
S(E_{i})({\rm keV.b})= \sum\limits_{\lambda=1}^{N}\,S_{\lambda}(E_{i})({\rm keV.b}),
\label{Sfctrsumlambda1}
\end{align}
\begin{widetext}
\begin{align}
S_{\lambda}(E_{i})({\rm keV.b})=
\nu_{N}^{2}\,E_{N}^{2}\,\frac{20\,\pi}{\mu_{i}}\,e^{2\,\pi\,\eta_{i}}\,\frac{ {\hat J}_{F^{(\lambda)}}    }{{\hat J}_{x}\,{\hat J}_{A}}\,  
 \frac{  P_{f_{\lambda}}\,P_{i_{\lambda}}\,\gamma_{f_{\lambda}}^{2}\,\gamma_{i_{\lambda}}^{2} }{ \Big(E_{\lambda} - E_{i}  - \sum\limits_{n=i,f}\,[S_{n\,\lambda}(E_{n}) - B_{n\,\lambda}) ]\,{\gamma}_{n\,\lambda}^{2}\Big)^{2} + \big[\sum\limits_{n=i,f}\,P_{n\,\lambda}\,\gamma_{n\,\lambda}^{2}\big] ^{2} }.  \nonumber\\
\label{Sfactrlambda2}
\end{align}   
\end{widetext}
Here, all the quantities with the subscripts $n,\lambda$  correspond to the channel $n$ and level $\lambda$, 
$\,\gamma_{i\,\lambda}$ and $\gamma_{f\,\lambda}$  are the reduced width amplitudes of the resonance $F^{(\lambda)}$   in the initial and final channels, $\gamma_{i\,1} \equiv \gamma_{i}^{(s)}$, $\,E_{\lambda}$ is the energy level in the channel $\lambda$.									

Now we consider  two interfering levels, $\lambda=1\,$ and $\,2$, and two channels in each level.  All the quantities related to the levels $\lambda=1$ and $2$ have additional subscripts $1$ or $2$, correspondingly.  We  assume that the level $\lambda=1$ corresponds to the subthreshold state  in the channel $i=x+A$, which decays to a resonant state corresponding to the level $\lambda=1$  in the channel $f=b+B$. The level $2$ describes the resonance in the channel $x+A$, which decays into the resonant state in the channel $f= b+ B$.  The level $\lambda =2$ lies higher than the level $\lambda=1$ but both levels do interfere.  
The reaction amplitude is given by
\begin{eqnarray}
T_{f\,i}=-2\,i\,e^{-i(\delta^{hs}_{f}+ \delta^{hs}_{i} )} \sqrt{P_{f}}\,\sqrt{ P_{i} }\sum\limits_{\lambda\,\tau}\gamma_{f_{\lambda}}\,{\rm {\bf A}}_{\lambda\,\tau}\,\gamma_{i_{\tau}}.
\label{M2lev2ch1}
\end{eqnarray}
Here, ${\rm {\bf A}}$   is the level matrix in the $R$-matrix method, 
\begin{eqnarray}
\big({\rm {\bf A}}^{-1}\big)_{\lambda\,\tau}&=& (-\varepsilon_{i}^{(s)} - E_{i} )\,\delta_{\lambda\,\tau} \nonumber\\
&-& \sum\limits_{n=i,f}\,\gamma_{n\,\lambda}\, \gamma_{n\,\tau}\,\big[S_{n}(E_{n}) - S_{n}(E_{n}^{(s)}) + i\,P_{n} \big]. \nonumber\\ 
\label{ARmatr1}
\end{eqnarray}

The corresponding astrophysical $S(E_{i})$ factor is
\begin{eqnarray}
S(E_{xA})({\rm keV.b})&=& 20\,\pi\,\nu_{N}^{2}\,E_{N}^{2}\frac{{\hat J}_{F^{(s)}}}{{\hat J}_{x}\,{\hat J}_{A}}\,\frac{1}{\mu_{i}}\,e^{2\,\pi\,\eta_{i}}\nonumber\\
&\times& P_{f}\,P_{i}\,  \Big|\sum\limits_{\lambda\,\tau}\,\gamma_{f_{\lambda}}\,{\rm {\bf A}}_{\lambda\,\tau}\,\gamma_{i_{\tau}}\,\Big|^{2}.
\label{Sfactr2l2ch1}
\end{eqnarray}

\section{Trojan Horse}

One of the most striking methods of treating low-energy resonances and subthreshold resonances is the Trojan Horse Method (THM), which is a powerful indirect technique allowing one to determine the astrophysical factor for rearrangement reactions \cite{reviewpaper}. While in  direct measurements, owing to the presence of the Coulomb barrier,  it is difficult or practically impossible to reach the region where the peak in the astrophysical factor is generated by the low-energy resonances or subthreshold resonance reveals itself, the THM is the only method, which allows one not only to observe the peak from these resonances at $E_{i} > 0$ but even to trace this peak down to the subthreshold bound state at $E_{xA}= - \varepsilon_{i}^{(s)}$.
The THM involves obtaining the cross section of the binary process (\ref{reaction1}) at astrophysical energies by measuring the Trojan Horse (TH) reaction [the two-body to three-body process ($2 \to 3$ particles)] in the quasi-free (QF) kinematics:
\begin{eqnarray}
a + A \to s + F^{*} \to s + b + B.
\label{THMreaction1}
\end{eqnarray}
The Trojan Horse particle, $a = (s\,x)$, which has a dominant $s$-wave cluster structure, is accelerated at energies above the Coulomb barrier. After penetrating the barrier, the TH-nucleus $a$ undergoes breakup leaving particle $x$ to interact with target $A$ while projectile $s$, also called a spectator, flies away. From the measured cross section of the TH reaction, the energy dependence of the astrophysical factor of the binary sub-process (\ref{reaction1}) is determined. Since the transferred particle $x$ in the TH reaction is virtual, its energy and momentum are not related by the On-the-Energy-Shell (OES) equation, that is, $\,E_{x}\not= k_{x}^{2}/(2\,m_{x})$. The main advantage of the THM is that the penetrability factor $P_{i}$  in the entrance channel of the binary reaction (\ref{reaction1}) is not present in the expression for the TH cross section \cite{Bau86}. It allows one to study resonant reactions  (\ref{reaction1})  at astrophysically relevant energies for which direct measurements are impossible or extremely difficult to perform  because of the very small  value of $\,P_{i}$.  The second advantage of  the THM  is that it provides a possibility to measure the cross section of the binary reaction (\ref{reaction1})  at negative $E_{i}$ owing to the off-shell character of the transferred particle $x$ in the TH reaction.

\begin{figure}[tbp] 
  \centering
  \includegraphics[scale=0.55,trim={180 240 0 3cm},clip]{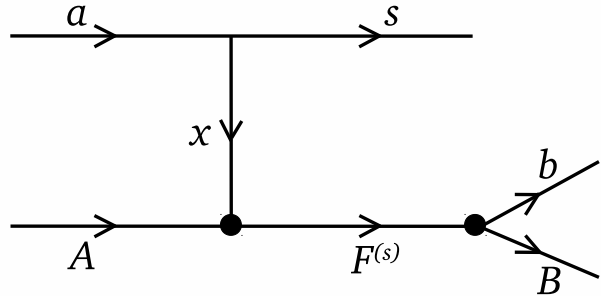}
  \caption{Pole diagram describing the TH reaction mechanism}
  \label{fig_THdiagram}
\end{figure}

\subsection{Trojan Horse reaction amplitude}

The details of the THM are addressed in the review paper \cite{reviewpaper}. The expression for the amplitude of the  reaction (\ref{THMreaction1}) ( for $x=n$), which is described by the diagram of Fig. 1,   in the surface integral approach and distorted wave Born approximation (DWBA) has been derived in \cite{muk2011}.  In the THM the absolute cross section of the reaction (\ref{THMreaction1}) is not measured and is determined by normalizing  the THM cross section to  available direct experimental data at higher energies. That is why it  makes  sense to use the plane wave approximation to get the THM amplitude. The expression for the prior form of the THM resonant reaction amplitude in the plane wave approximation  was derived in \cite{reviewpaper} using the generalized $R$-matrix approach for the three-body reactions. The derived expression is valid for the reactions proceeding through the real  and subthreshold resonances and
takes the form (with fixed projections of the spins of the initial and final particles)
\begin{widetext}
\begin{align}
& M_{M_{a}\,M_{A}}^{M_{s}\,M_{b}\,M_{B}}= i\,2\,{\pi}^{2}\,\sqrt{\frac{1}{\mu_{f}\,k_{f}}}\,\varphi_{a}( p_{sx})\, \sum\limits_{J_{F^{(s)}}\,l_{f}\,l_{i}}\,i^{l_{f}+l_{i}}
\sum\limits_{M_{x}\,M_{F^{(s)}}\,m_{j_{i}} \, m_{j_{f}}\,m_{l_{f}} \, m_{l_{i}}}  e^{-i\,\delta_{f}^{hs}}\,Y_{l_{f}\,m_{l_{f}}}(-{\rm {\bf {\hat  k}}}_{f})\,\sqrt{\frac{R_{i}}{\mu_{i}}}e^{-i\,\delta_{i}^{(hs)}}\,\nonumber\\
&\times Y_{l_{i}\,m_{l_{i}}}^{*}({\rm {\bf {\hat p}}}_{i})\,\langle j_{f}\,m_{j_{f}}\,\,l_{f}\,m_{l_{f}}|J_{F^{(s)}}\,M_{F^{(s)}} \rangle \langle j_{i}\,m_{j_{i}}\,\,l_{i}\,m_{l_{i}}|J_{F^{(s)}}\,M_{F^{(s)}}\rangle\, \langle J_{b}\,M_{b}\,\,J_{B}\,M_{B}| j_{f}\,m_{j_{f}} \rangle\,\langle J_{x}\,M_{x}\,\,J_{A}\,M_{A}|j_{i}\,m_{j_{i}}\rangle \,\nonumber\\
&\times \langle J_{s}\,M_{s}\,\,J_{x}\,M_{x}| J_{a}\,M_{a} \rangle \, P^{-1/2}_{l_{i}}\,T_{l_{f}\,l_{i}}\,{\tilde M}_{l_{i}}.
\label{THM1l2ch1}
\end{align}
\end{widetext}
Here
\begin{align}
{\tilde M}_{l_{i}} &= j_{l_{i}}(p_{i}\,R_{i})\Bigg[[B_{i}-1] -D_{i}\Bigg] \nonumber\\
&+  \frac{ 2Z_{x}\,Z_{A}\,\mu_{xA}}{137}\int\limits_{R_{i}}^{\infty} {\rm d}r_{i} \frac{O_{l_{i}}(k_{i},\,r_{i})}{  O_{l_{i}} (k_{i},R_{i}) }\,j_{l_{i}}(p_{i},\,r_{i}),
\label{Mli1}
\end{align}
with
\begin{eqnarray}
D_{i} \equiv D_{l_{i}}(p_{i},\,R_{i})= R_{i}\,\frac{{\partial j_{l_{i}}(p_{i},r_{i})}/{\partial r_{i}}\Big|_{r_{i}=R_{i}}}{j_{l_{i}}(p_{i},R_{i})}, \nonumber\\
B_{i} \equiv B_{l_{i}}(k_{i},\,R_{i})= R_{i}\,\frac{{\partial O_{l_{i}}(k_{i},r_{i})}/{\partial r_{i}}\Big|_{r_{i}=R_{i}}}{O_{l_{i}}(k_{i},R_{i})}.
\label{DB1}
\end{eqnarray}
Note that $i=x+A$ and $f=b+B$ determine the entry and exit channels of the subreaction (\ref{reaction1})  rather than the entry and exit channels of the TH reaction  (\ref{THMreaction1}).
In Eq. (\ref{THM1l2ch1})  $\,<j_{1}\,m_{j_{1}}\,\,j_{2}\,m_{j_{2}}\big| j\,m_{j}>$ is the Clebsch-Gordan coefficient,   $\,J_{j} (M_{j})$  is the spin (its projection) of particle $j$, $\,j_{n} (m_{j_{n}})$ is the channel spin (its projection) in the channel $\,n=i,\,f$ and $\,m_{l_{n}}$ is the projection of $\,l_{n}$; the sum over projection $\,M_{J_{F^{(s)}}}\,$ of the spin $\,J_{F^{(s)}}\,$ is added formally and can be dropped because the projections of the spins of the final particles $b$ and $B$ are fixed.  Also, $\,\varphi_{a}( p_{sx})$ is the Fourier transform of the $s$-wave bound-state wave function of nucleus $a$; because we included the sum over $l_{i}$  we replaced  $\,P_{i}\,$ by  $P_{l_{i}}$;
$\,\,O_{l_{i}}(k_{i},\,r_{i})$ is the outgoing spherical wave, $j_{l_{i}}(p_{i},R_{i})$ is the spherical Bessel function. We also remind that
$r_{i} \equiv r_{xA}$ is the radius connecting $\,x$ and $\,A$, $k_{i}=k_{xA}$ is the conjugated OES $\,x-A$ relative momentum, that is, $\,E_{i} \equiv E_{xA} = 2\,k_{i}^{2}/(2\,\mu_{i})$, $\,\mu_{i}= \mu_{xA}$ is the reduced mass of particles $x$ and $A$,
${\rm {\bf k}}_{f} \equiv {\rm {\bf k}}_{bB}$.
 $\,{\rm {\bf p}}_{i} \equiv {\rm {\bf p}}_{xA}$ and $ {\rm {\bf p}}_{sx}$  are  the off-shell $x-A$ and $s-x$ relative momenta, correspondingly. 

We see that the THM reaction amplitude is expressed in terms of the  binary sub-reaction OES amplitude $T_{l_{f}\,l_{i}}$ in the $R$-matrix form. We slightly modified the notation of the binary sub-reaction amplitude adding the subscripts $l_{i}$ and $l_{f}$ to underscore explicitly a dependence of the binary reaction amplitude on the $x-A$ and $b-B$ relative orbital angular momenta
in the entry channel and the exit channels, correspondingly, of the reaction (\ref{reaction1}).  $T_{l_{f}\,l_{i}}$ is given by any of the resonance  amplitudes derived above. 

 The possibility to express the TH reaction amplitude in terms of the $R$-matrix OES amplitude $T_{l_{f}\,l_{i}}$ is the result of  using the surface integral formalism
and the generalized $R$-matrix approach for the three-body reactions \cite{muk2011,reviewpaper}. The most important feature of the THM providing the success of the THM as an indirect technique for the analysis of the low-energy resonances, which are not reachable by direct measurements, is the absence of the penetrability factor $P_{l_{i}}$ in the entry channels of the binary sub-reaction
(\ref{reaction1}). The absence of $P_{l_{i}}$ is evident because $P_{l_{i}}$ containing in $T_{l_{f}\,l_{i}}$ is compensated by the factor $P_{l_{i}}^{-1}$ in front of   $T_{l_{f}\,l_{i}}$.
  
However, we pay a price for using the three-body reaction to obtain an information about the two-body sub-reaction.  Two additional factor appears in the TH reaction amplitude. The first factor is $\,{\tilde M}_{l_{i}}$. This factor contains the logarithmic derivatives of the spherical Bessel function $\,j_{l_{i}}(p_{i},r_{i})$ and the outgoing spherical wave $\,O_{l_{i}}(k_{i},\,r_{i})$ taken at the channel radius $\,R_{i}$. These logarithmic derivatives are a result of the generalized $R$-matrix approach. In addition, $\,{\tilde M}_{l_{i}}$ contains the radial integral whose  integrand behaves asymptotically (at $r_{i} \to \infty$) as 
$ \sim  {\sin(p_{xA}\,r_{i} -l_{i}\pi/2)}({p_{xA}\,r_{i}})^{-1}r_{i}^{-i\eta_{xA}}$. Such an integral is not converging in the usual sense and requires a regularization to provide convergence at large $r_{i}$. This integral was not taken into account in previous THM publications. 
Another important factor appearing from the consideration of the three-body TH reaction is the Fourier transform of the bound-state wave function of the TH particle $\varphi_{a}( p_{sx})$, which plays an important role in the determination of TH reaction kinematics. As we have underscored, in the THM the selected loosely-bound TH particle $a = (s\,x)$ has a dominant $s$-wave cluster structure. It is necessary for the following reasons:\\ 
(i) The Fourier transform of the $s$-wave bound-state wave function has a peak at $p_{sx}=0$. The reaction kinematics for which $p_{sx}=0$ is called quasi-free (QF) one. In the practical application to cover some energy interval $E_{xA}$ at fixed $E_{aA}$
 one needs to select the THM reaction events with different $p_{sx}$. The larger $p_{sx}$ the smaller $\varphi_{a}( p_{sx})$ and, hence, the smaller the TH cross section. But for a loosely-bound TH particle $a = (s\,x)$ the decay of $\varphi_{a}( p_{sx})$ with increase of $p_{sx}$ is much smaller than for strongly-bound particles. Typically in the THM  a variation of $p_{sx}$ is carried out in the interval $p_{sx} \leq \kappa_{sx}$ where $\kappa_{sx}=\sqrt{2\,\mu_{sx}\,\varepsilon_{sx}}$ is the TH particle wave number, $\varepsilon_{sx}$ is the binding energy of $a$ with respect to the virtual decay $a \to s+ x$ and $\mu_{s\,x}$ is the reduced mass of $s$ and $x$. Hence, in the THM with loosely bound TH particles one has the priviledge to deviate from the QF kinematics within the interval $p_{sx} \leq \kappa_{sx}$ without loosing significantly the value of the TH cross section. \\
(ii) There is another important reason of choosing a loosely-bound TH particle. At $p_{sx} \leq \kappa_{sx}$ the probable distances between $s$ and $x$ are $r_{sx} \geq 1/\kappa_{sx}$. The smaller $\kappa_{sx}$ the larger $r_{sx}$. At large $r_{sx}$ one can treat the outgoing particle $s$ as a spectator which causes a minimal disturbance to the binary sub-reaction (\ref{reaction1}). 

The factor ${\tilde M}_{l_{i}}$ depends on the relative $x-A$ momentum $p_{xA}$.  From momentum conservation we get (see diagram in 
Fig. \ref{fig_THdiagram})
\begin{align}
{\rm {\bf p}}_{xA}= \frac{m_{A}\,{\rm {\bf p}}_{x} - m_{x}\,{\rm {\bf k}}_{A}}{m_{xA} },
\label{pxA1}
\end{align}
where ${\rm {\bf p}}_{x}= {\rm {\bf k}}_{a} - {\rm {\bf k}}_{s}$ is the momentum of the virtual particle $x$ and ${\rm {\bf k}}_{j}$ is the momentum of the real particle $j$, $m_{ij}= m_{i} + m_{j}$.  It is convenient to consider the system in which the TH particle $a$ is at rest, that is, ${\rm {\bf k}}_{a}= {\rm {\bf k}}_{s} + {\rm {\bf p}}_{x}=0$. Then  
\begin{align}
 {\rm {\bf p}}_{xA}= -\frac{m_{A}\,{\rm {\bf k}}_{s} + m_{x}\,{\rm {\bf k}}_{A}}{m_{xA} }
\label{pxA2}
\end{align}
and ${\rm {\bf p}}_{sx}= {\rm {\bf k}}_{s}$. 

There is another THM important relation: the particle $x$ is virtual. Hence, $E_{x} - p_{x}^{2}/(2\,m_{x}) \not=0$. From the momentum-energy conservation in the three-ray vertices $a \to s+ x$ and $x+A \to F$ in the diagram in Fig. \ref{fig_THdiagram}
we get
\begin{align}
E_{xA}= \frac{p_{xA}^{2}}{2\,\mu_{xA}} - \frac{p_{sx}^{2}}{2\,\mu_{sx}} - \varepsilon_{sx}.
\label{ExA1}
\end{align}
From this equation we can conclude that always ${p_{xA}^{2}}/{2\,\mu_{xA}}>E_{xA}$. Moreover, the binding energy of the TH particle $a$ plays an important role in decreasing $E_{xA}$ allowing one to measure the TH cross section at lower energies even if the beam energy is higher than the Coulomb barrier in the initial channel $a+A$ of the TH reaction (\ref{THMreaction1}) \cite{Bau86,reviewpaper}.
It is convenient to rewrite Eq. (\ref{ExA1}) in the system where $k_{a}=0$. In this system Eq. (\ref{ExA1}) can be reduced to
\begin{align}
E_{xA}= \frac{m_{x}}{m_{xA}}\,E_{A} + \frac{{\rm {\bf k}}_{s} \cdot {\rm {\bf k}}_{A}}{m_{xA}} - \frac{k_{s}^{2}}{2\,\mu_{sx}} - \varepsilon_{sx}.
\label{ExA2}
\end{align}
 
\subsection{Triple and double differential cross sections of Trojan Horse reaction}

The TH triple differential cross section is given by
\begin{align}
\frac{{\rm d}^{3}\sigma}{{\rm d}\Omega_{{\rm {\bf k}}_{bB}}\,{\rm d}\Omega_{{\rm {\bf k}}_{sF}}\,{\rm d}E_{xA}} &= \frac{\mu_{aA}\,\mu_{sF}\,\mu_{bB}}{ (2\,\pi)^{5}}\,\frac{k_{sF}\,k_{bB}}{k_{aA}}\, \frac{1}{ {\hat J}_{a}\,{\hat J}_{A}  }\nonumber\\
&\times \sum\limits_{M_{a}\,M_{A}\,M_{s}\,M_{b}\,M_{B}}\,\big|
M_{M_{a}\,M_{A}}^{M_{s}\,M_{b}\,M_{B}} \big|^{2}.
\label{3diffcrsect1}
\end{align}
We remind that $E_{bB}$ and $E_{xA}$ are related by Eq. (\ref{EfEiQ1}).  That is why we replaced ${\rm d}E_{bB}$ by ${\rm d}E_{xA}$. 
For practical applications it is more convenient to use the TH double differential cross section,
which is obtained by integration of the the triple differential cross section over $\Omega_{{\rm {\bf k}}_{bB}}$.
Using the orthogonality of the Clebsch-Gordan coefficients and integrating over $\Omega_{{\rm {\bf k}}_{bB}}$ we get for the TH double differential cross section
\begin{align}
\frac{{\rm d}^{2}\sigma}{{\rm d}\Omega_{{\rm {\bf k}}_{sF}}\,{\rm d}E_{xA}} &=  \frac{ e^{-2\,\pi\,\eta_{xA}} \varphi_{a}^{2}(p_{sx}) }{160\,{\pi}^{3}\,\nu_{N}^{2}\,E_{N}^{2}}\,\,R_{xA}\,\mu_{aA}\,\mu_{sF}\,\frac{k_{sF}}{k_{aA}}\nonumber\\
&\times  \sum\limits_{l_{i}}\,P_{l_{i}}^{-1}\,S_{l_{i}}(E_{xA})\Big|{\tilde M}_{l_{i}}\Big|^{2}.
\label{doubledifcrsect1}
\end{align}
We remind that $l_{i} \equiv l_{xA}$. 
Thus the TH double differential cross section is expressed in terms of the OES astrophysical factors $S_{l_{i}}(E_{xA})$.  By measuring the energy dependence of the TH double differential cross section we actually measure the energy dependence of the OES astrophysical factor. We underscore again that in the THM only the energy dependence of the double differential cross section on $E_{xA}$ is measured. 

To extract the astrophysical factor from the TH experiment we assume, for simplicity, that in the region where direct data are available only one $l_{i}$ gives a dominant contribution.  Then expressing the astrophysical factor in terms of the TH double differential cross section and introducing normalization factor of the TH astrophysical factor to the available experimental data  at higher energies, at which penetrability factor $P_{l_{i}}$ is not an issue and direct measurements are available,  we get  the TH astrophysical factor
\begin{align}
S_{l_{i}}(E_{xA})&= NF\, e^{2\,\pi\,\eta_{xA}} \frac{k_{aA}}{k_{sF}}\,\frac{160\,{\pi}^{3}\,\nu_{N}^{2}\,E_{N}^{2}}{R_{xA}\,\mu_{aA}\,\mu_{sF}} \nonumber\\
&\times P_{l_{i}}\,\frac{1}{\varphi_{a}^{2}(p_{sx})\,\Big|{\tilde M}_{l_{i}}\Big|^{2} }\,\,\frac{{\rm d}^{2}\sigma}{{\rm d}\Omega_{{\rm {\bf k}}_{sF}}\,{\rm d}E_{xA}}
\label{THastrfactorNorm1}.
\end{align}
Here, $NF$ is the energy-independent TH normalization factor. After the NF has been determined at higher energy one can determine the astrophysical factor at lower energies using the experimental 
TH double differential cross section. 

Assume that at low energies only one $l_{i}$ does contribute then Eq. (\ref{THastrfactorNorm1}) can be used to determine $S_{l_{i}}$ at lower energies. If there are two or more interfering resonances then all of them have the same $l_{i}$.  If, for example, two  resonances contribute with different $l_{i}$ then one can find a region where one of these resonances dominate. Once the astrophysical factor for one of the resonances is determined, the astrophysical factor for the second one can be also determined.

\section{Astrophysical factor  of  $^{13}{\rm C}(\alpha,\,n)^{16}{\rm O}$ reaction}

The $^{13}{\rm C}(\alpha,\,n)^{16}{\rm O}$ reaction is considered to be the main neutron supply to build up heavy elements from iron-peak seed nuclei in AGB stars. At temperature $0.9 \times 10^{8}$ K, the energy range where the $^{13}{\rm C}(\alpha,\,n)^{16}{\rm O}$ reaction is most effective, the so-called Gamow window \cite{illiadis,BK16} is within $\approx 140 - 230$ keV with the most effective energy at $\approx 190$ keV.  This reaction was studied using both direct and indirect (TH) methods.  Direct data, owing to the small penetrability factor, were measured with reasonable accuracy down to $E_{\alpha\,{}^{13}{\rm C}} \approx 400$ keV. Data in the interval $300 - 400 $ keV were obtained with much larger uncertainty \cite{davids68,bairhaas,drotleff,brune,Harissopulos}.  In the paper \cite{Harissopulos}  the unprecedented accuracy of $4\%$  was achieved at energies $E_{\alpha\,{}^{13}{\rm C}}> 600$ keV.  The dominant contribution to the $^{13}{\rm C}(\alpha,\,n)^{16}{\rm O}$  reaction at astrophysical energies comes from the  state ${}^{17}{\rm O}(1/2^{+}, E_{x}=6356 \pm 8 \, {\rm keV})$, where $E_{x}$ is the excitation energy. Taking into account that the $\alpha- {}^{13}{\rm C}$ threshold is located at $6359.2$ keV one finds that this $1/2^{+}$ level is the located at $E_{\alpha\,{}^{13}{\rm C}} = -3 \pm 8$ keV, that is, it can be or subthreshold bound state or a resonance \cite{tiley}. This location of the level ${}^{17}{\rm O}(1/2^{+})$ was adopted in the previous analyses of the direct measurements including the latest one in \cite{heil}. If this level is the subthreshold bound state, then its reduced width is related to ANC of this level.

 However,  in the recent paper \cite{Faestermann} it has been determined that this level is actually a resonance located at $E_{\alpha\,{}^{13}{\rm C}} = 4.7 \pm 3$ keV with the total observable width  of ${\tilde \Gamma} =136 \pm 5 $ keV.  Note that ${\tilde \Gamma}_{\alpha}$ of this resonance with $l_{i}=1$ is negligibly small because it contains the penetrability factor $P_{1}$.  Hence, ${\tilde \Gamma}= {\tilde \Gamma}_{n}$.
The result obtained in \cite{Faestermann} is a very important achievement in the long history of hunting for this near threshold level. If this level is actually a resonance located slightly above the threshold then the reduced width is related to the resonance partial $\alpha$ width rather than to  the  ANC. Evidently that this resonance is not a Breit-Wigner type and it does not make sense to use the ANC as characteristics of this resonance (see Eq. (\ref{ANCGammawideres1})).

Here we present the calculations of the astrophysical $S$-factors for the $^{13}{\rm C}(\alpha,\,n)^{16}{\rm O}$ using the  equations derived above.   We fit the latest TH data \cite{oscar}  using  both assumptions that the threshold level $1/2^{+}$  is the subthreshold state  located at $-3$ keV and  the resonance state at $4.7$ keV. For the subthreshold state we use parameters from \cite{heil} while for the resonance state we adopted parameters from \cite{Faestermann}. The resonances included in the analysis of this reaction are ($\,1/2^{+},\,l_{i}=1,\,E_{x}=6.356$ MeV), ($\,5/2^{-},\,l_{i}=2,\,E_{x}=7.165$ MeV), ($\,3/2^{+},\,l_{i}=1,\,E_{x}=7.216$ MeV), ($5/2^{+},\,l_{i}=3,\,E_{x}=7.379$ MeV) and ($5/2^{-},\,l_{i}=2,\,E_{x}=7.382$ MeV). Only two resonances, the second and the last one have the same quantum numbers and do interfere. Their interference can be taken into account  using  the S-factor given by Eq. (\ref{Sfactr2l2ch1}). For non-interfering resonances we  use Eq. (\ref{Sfctrsumlambda1}). 

\begin{figure}[ht]
  \includegraphics[height=7.5cm, clip,width=0.45\textwidth]{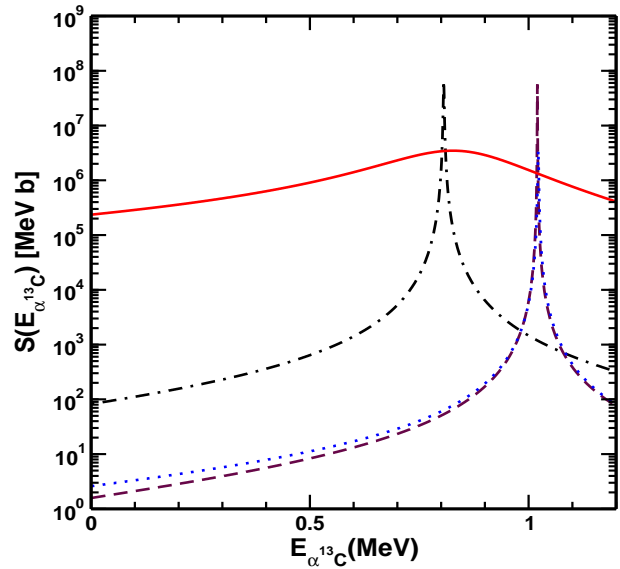}
\caption{(color online) 
The $S$-factors for the $^{13}{\rm C}(\alpha,\,n)^{16}{\rm O}$ reaction as a function of the $\alpha-{}^{13}{\rm C}$ relative kinetic energy
proceeding through four resonances: black dotted-dashed line -($\,5/2^{-},\,l_{i}=2,\,E_{x}=7.165$ MeV; solid red line- ($\,3/2^{+},\,l_{i}=1,\,E_{x}=7.216$ MeV); dashed brown line-($5/2^{+},\,l_{i}=3,\,E_{x}=7.379$ MeV); dotted blue line- ($5/2^{-},\,l_{i}=2,\,E_{x}=7.382$ MeV). All the resonant parameters are taken from \cite{heil}.}  
\label{fig_energdepres1}
\end{figure}

In Fig. \ref{fig_energdepres1} we presented the $S$ factors  contributed by four different resonant states located at $E_{\alpha\,{}^{13}{\rm C}}>0$.  All the parameters of these resonances are taken from \cite{heil}. We only slightly modified the $\alpha$-particle width of the wide resonance at $E_{\alpha\,{}^{13}{\rm C}}=0.857$ MeV taking it to be $0.12$ keV. The adopted channel radii are 
$R_{\alpha\,{}^{13}{\rm C}}= 7.5$ fm and $R_{n\,{}^{16}{\rm O}}= 6.0$ fm.

As we see from  Fig. \ref{fig_energdepres1}, the contributions of  all the narrow resonances are negligible compared to the wide one (red solid line in Fig. \ref{fig_energdepres1}). That is why we do not take into account the interference between two narrow $5/2^{-}$ resonances. 
Thus eventually we can take into account only the wide resonance ($\,3/2^{+},\,l_{i}=1,\,E_{x}=7.216$ MeV) and the near threshold level ($\,1/2^{+},\,l_{i}=1,\,E_{x}=6.356$ MeV).

\subsection{Threshold level  $\,1/2^{+},\,l_{i}=1,\,E_{x}=6.356$ MeV}

Here we would like to discuss the threshold  level $E_{x}=6.356$ MeV. Until the work of Ref. \cite{Faestermann} this level was considered to be the subthreshold resonance located at $E_{\alpha\,{}^{13}{\rm C}} = -3$ keV.
However, as we have mentioned,  in \cite{Faestermann} this level now was shifted to the continuum and is found to be a real resonance located at $E_{\alpha\,{}^{13}{\rm C}}= 4.7$ keV. The astrophysical factor contributed by this $1/2^{+}$ state depends on the  reduced width in the entry channel $\alpha-{}^{13}{\rm C}$ of the $^{13}{\rm C}(\alpha,\,n)^{16}{\rm O}$ reaction and the reduced width in the exit channel $n-{}^{16}{\rm O}$. The latter is determined with an acceptable accuracy, for example, in \cite{heil,Faestermann}.
If we assume that the level $E_{x}=6.356$ MeV is the subthreshold resonance then its reduced width in the $\alpha$-channel is expressed in terms of the ANC for the virtual decay ${}^{17}{\rm O}(1/2^{+},E_{\alpha\,{}^{13}{\rm C}} = -3\,\,{\rm keV} )
\to \alpha + {}^{13}{\rm C}$. This ANC was found in a few papers \cite{pellegriti,guo, marco2}. The latest measurement of this ANC was published in \cite{avila}: ${\tilde C}_{\alpha\,{}^{13}{\rm C}}^{(s)}= 3.6 \pm 0.7$ fm${}^{-1}$, which is the Coulomb renormalized ANC. The problem is that at very small binding energies the ANC of the subthreshold state becomes very large due to the Coulomb-centrifugal barrier. That is why in \cite{echaya,muk2012}  the Coulomb renormalized ANC was introduced in which the Coulomb-centrifugal factor was removed:
\begin{align}
{\tilde C}= \frac{l_{i}!}{\Gamma(1+ l_{i}+ \eta^{(s)})}\,C.
\label{ANCCoulren1}
\end{align} 
Here, $\,\Gamma(x)\,$ is the Gamma-function, $\,l_{i}\,$ is the orbital angular momentum of the bound state, $\,\eta^{(s)}\,$ is the Coulomb parameter of the subthreshold bound state. At small binding energies of the bound state, that is, at large $\,\eta^{(s)}$, the factor $\,\Gamma(1+ l_{i}+ \eta^{(s)})\,$ becomes huge. Usually we are used to see that the barrier factor decreases the cross section but here we see the opposite effect.  

However, in the $R$-matrix approach the quantity, which we need to calculate the astrophysical $S$-factor, is the reduced width. The observable reduced width of the bound state is expressed  in terms of the ANC by equation 
\begin{align}  
{\tilde \gamma}^{2}= \frac{C^{2}\,W_{-\,\eta^{(s)},\,l_{i}+1/2}^{2}(2\,\kappa_{i}^{(s)}\,R_{i}
)}{2\,\mu\,R_{i}}.
\label{redwidthANC2}
\end{align}
 The Coulomb-barrier factor, which significantly enhances the ANC,  makes an opposite effect on the Whittaker function $W_{-\,\eta^{(s)},\,l_{i}+1/2}(2\,\kappa_{i}^{(s)}\,R_{i})$, so that the product
$C\,W_{-\,\eta^{(s)},\,l_{i}+1/2}(2\,\kappa^{(s)}\,R)$ is unaffected by the Coulomb-centrifugal barrier factor.  It is convenient to rewrite 
Eq. (\ref{redwidthANC2}) as
\begin{align}  
{\tilde \gamma}^{2}= \frac{{\tilde C}^{2}\,{\tilde W}_{-\,\eta^{(s)},\,l_{i}+1/2}^{2}(2\,\kappa_{i}^{(s)}\,R_{i})}{2\,\mu_{i}\,R_{i}},
\label{redwidthANC3}
\end{align}
where $R_{i}= R_{\alpha\,{}^{13}{\rm C}}$,  $\mu_{i}=\mu_{\alpha\,{}^{13}{\rm C}}$, $\kappa_{i}^{(s)}= \sqrt{2\,\mu_{i}\varepsilon_{i}^{(s)}}$,
$\varepsilon_{i}^{(s)}=-3$  keV. Also
\begin{align}
{\tilde W}_{-\,\eta^{(s)},\,l_{i}+1/2}^{2}(2\,\kappa^{(s)}\,R_{i}) &= \frac{\Gamma(1+ l_{i}+ \eta^{(s)})}{l_{i}!}\nonumber\\
&\times W_{-\,\eta^{(s)},\,l_{i}+1/2}^{2}(2\,\kappa^{(s)}\,R_{i}).
\label{tildeW1}
\end{align}
For example, for the case under consideration, if the subthreshold bound state is located at $-3$ keV then
$\Gamma(1+ l_{i}+ \eta^{(s)})=2.406 \times 10^{84}$ for $l_{i}=1$. For the channel radius $R_{i}=7.5$ fm,  
$W_{-\,\eta^{(s)},\,l_{i}+1/2}(2\,\kappa^{(s)}\,R_{i})=2.44122 \times 10^{-86}$ while ${\tilde W}_{-\,\eta^{(s)},\,l_{i}+1/2}(2\,\kappa_{i}^{(s)}\,R_{i})= 0.0587$.  Correspondingly, 
\begin{align}
C\,W_{-\,\eta^{(s)},\,l+1/2}(2\,\kappa_{i}^{(s)}\,R_{i})&= {\tilde C}\,{\tilde W}_{-\,\eta^{(s)},\,l+1/2}(2\,\kappa_{i}^{(s)}\,R_{i})\nonumber\\
&= 0.111 \, {\rm fm}{}^{-1/2}.
\label{CWtildeCW1}
\end{align}

The reduced width changes very little if we assume that the threshold level $1/2^{+}$ is the bound state. We used the single-particle $\alpha-^{13}{\rm C}$
Woods-Saxon potential to generate the bound-state wave function with the binding energy $-3$ keV.  This function has three nodes at $r_{i}>0$. Following the $R$-matrix procedure, we accepted the internal region as $0 \leq r \leq R$, where $R=5.2$ fm is the location of the last peak of the internal wave function, and calculated the wave function, which is normalized over the internal region, at  $R=5.2$ fm. The obtained value can give estimation of the single-particle reduced width amplitude.  
After that we adopted the binding energy as $-0.1$ keV and repeated the similar procedure and found by decreasing the well-depth that $R=4.93$ fm. The  value of the single-particle reduced width decreased only by $2.5\%$ compared to the value for the binding energy of $-3$ keV. Because the reduced width of the 
resonance state at $4.7$ keV is unknown and we are not able to reproduce this state using a single-particle Woods-Saxon potential, as we did for the bound states, 
we assume that the reduced width for the resonance state is close to the reduced width for the bound state $-3$ keV, which is $3.3$ keV${}^{1/2}$ for the ANC
${\tilde C}=1.9$ fm${}^{-1/2}$ and $R_{i}=7.5$ fm. To make the fit  to the TH data \cite{oscar} we adopted the reduced width for the resonance state $4.7$ keV  in the interval $\big(2.81-3.6 \big)$ keV${}^{1/2}$. Note that  $R_{i}= 7.5$ fm provided the best fit of the TH data.

\subsection{Low-energy astrophysical factor for ${}^{13}{\rm C}(\alpha,\,n){}^{16}{\rm O}$}

\begin{figure}[ht]
\includegraphics[height=7.5cm, clip,width=0.45\textwidth]{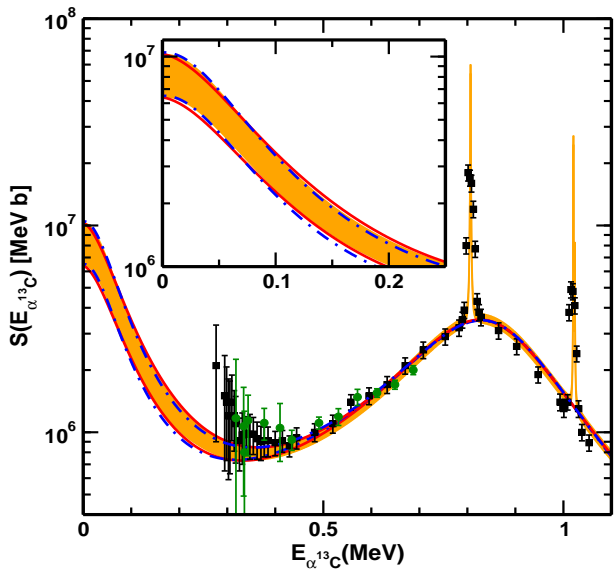}
\caption{(color online) Astrophysical S-factor for the $^{13}{\rm C}(\alpha,\,n)^{16}{\rm O}$ reaction as a function of the $\alpha\,{}^{13}{\rm C}$ relative kinetic energy. Square black boxes, solid green dots and shaded orange band are data from Refs. \cite{drotleff}, \cite{heil} and \cite{oscar}, respectively. Red solid lines correspond to our calculations for the fit to the lower and upper limits of the TH data  considering  $1/2^{+}$ state as  $-3$ keV subthreshold  resonance with $\Gamma_{n} = 158.1$ keV \cite{heil}. The lower and upper limits of ANC square are 2.89 fm$^{-1}$ and 4.7 fm$^{-1}$, respectively. Whereas, the blue dotted-dashed lines correspond to our calculations for the fit to TH data, considering  $1/2^{+}$ state as 4.7 keV threshold resonance with $\Gamma_{n} = 136$ keV \cite{Faestermann} and the corresponding lower and upper values of observable reduced width are 2.81 keV$^{1/2}$ and 3.6 keV$^{1/2}$, respectively.  For our calculations we have used $R_{\alpha\,{}^{13}{\rm C}}= 7.5$ fm, $R_{n\,{}^{16}{\rm O}}= 6.0$ fm. The insert in the figure shows enlarged low energy $S$-factor.  } 
\label{fig_Sfactor}
\end{figure}
From Fig. \ref{fig_energdepres1} it is clear that only the experimental $S$-factor generated by the broad resonance   $3/2^{+},\,E_{\alpha\,{}^{13}{\rm C}}=0.857$ MeV can be used for normalization of the TH double differential cross section at $E_{\alpha\,{}^{13}{\rm C}} > 0.5$ MeV. The problem of the normalization of the TH data for this specific reaction was discussed in details in \cite{marco2,oscar}.
We use the results from \cite{oscar} as fitting data but need to renormalize them because in \cite{marco2,oscar} the factor ${\tilde M}_{1}$  was calculated without the integral term in Eq. (\ref{Mli1}). Recalculating  ${\tilde M}_{1}$  taking into account the integral term we find that the TH results in \cite{oscar} should be renormalized by $0.948$. After renormalization of the TH data from \cite{oscar} we did a new fit. In Fig. \ref{fig_Sfactor} we present our final results for the $S$ factor for the reaction  ${}^{13}{\rm C}(\alpha,\,n){}^{16}{\rm O}$.

Our numerical values of  the $S(0)$ factors are:\\
(1) for  $1/2^{+}$,  $-3$ keV and $\Gamma_{n}= 158.1$ keV  \cite{heil},  $S(0)=7.62 _{-1.23}^{+2.65}\times 10^{6}$ MeV.b;\\
(2) for  $1/2^{+}$,   4.7 keV   and $\Gamma_{n}= 136$ keV \cite{Faestermann}, $S(0)=7.51 _{-1.1}^{+2.96}\times 10^{6}$ MeV.b. 

Thus, even the TH data, which  provides the astrophysical factor at significantly lower energies than direct measurements \cite{heil}, cannot  answer the question whether the threshold level  is  a subthreshold bound state or resonance.

In the analysis of the TH data, only the two-stage mechanism proceeding through the intermediate threshold state $1/2^+$ has been taken into account in this paper and in the previous TH papers (see Refs. \cite{oscar,marco2}). However, the single-step direct reaction $^{13}$C($\alpha, n$)$^{16}$O also can contribute to the low-energy cross section. Although the S-factor of the direct mechanism is flat and can be small its interference with the two-stage resonant mechanism can change the total S-factor. However, the accuracy of the existing data does not allow us to determine the contribution of the direct mechanism.

\section{Summary}
In this paper we discussed the $R$-matrix approach to the subthreshold resonances for the single-level and one channel,  and for the single-level and two channel cases. The connection between the  observable reduced width and the ANC is presented for the single-level, single-channel case and generalized for the two-channel case. We present the relationship between the resonance width and the ANC for general case and consider two limiting cases: broad and narrow resonances.  It is demonstrated how the resonant reactions proceeding through the subthreshold resonance can be treated within the conventional $R$-matrix approach.

Different equations for the astrophysical factors in the $R$-matrix approach are presented, which we use to calculate the astrophysical factor for the 
${}^{13}{\rm C}(\alpha,n){}^{16}{\rm O}$. All the equations are written in the convenient forms which can be directly used by the readers.
Special attention is given to the THM formalism. Our equation for the TH amplitude  is obtained using the surface-integral formalism and generalized $R$-matrix approach for the three-body resonant reactions.  It is shown how the TH double differential cross section can be expressed in terms of the on-the-energy-shell astrophysical factor for the binary sub-reaction. 

Finally, we demonstrated how the THM method can be used to calculate the astrophysical factor for the neutron generator $^{13}{\rm C}(\alpha,\,n)^{16}{\rm O}$ in low-mass AGB stars. At astrophysically relevant energies this astrophysical factor is controlled by the threshold level $1/2^{+}, E_{x}=6356 $ keV.   Here, we reanalyzed  recent TH data \cite{oscar} using both assumptions that the threshold level is the subthreshold state and  that it is a resonance state.

\begin{acknowledgements}
A.M.M. acknowledges support from the U.S. DOE grant numbers DE-FG02-93ER40773 and DE-FG52-09NA29467 and by the U.S. NSF grant number PHY-1415656. C.A.B. acknowledges support from the U.S. NSF Grant number 1415656 and the U.S. DOE Grant number DE-FG02-08ER41533. We also thank  M. La Cognata and O.  Tippella for providing us  direct as well as indirect TH experimental data and for several useful correspondences.
\end{acknowledgements}

\end{document}